# Bunch width versus macrostep height: A quantitative study of the effects of step-step repulsion

Hristina Popova[1,*], Filip Krzyżewski[2], Magdalena Załuska-Kotur[2] and Vesselin Tonchev[3]

[1]*Institute of Physical Chemistry, Bulgarian Academy of Sciences, Acad. G. Bonchev Str., block 11, 1113 Sofia, Bulgaria*

[2]*Institute of Physics, Polish Academy of Sciences, al. Lotników 32/46, 02-668 Warsaw, Poland*

[3]*Faculty of Physics, Sofia University, 1164 Sofia, Bulgaria*

[*]Corresponding author: karleva@ipc.bas.bg

**Abstract**

Bunching of steps at the surface of growing crystals can be induced by both directions of the driving force: step up and step down. The processes happen in different adatom concentrations and differ in character. In this study we show how the overall picture of the bunching process depends on the strength of short range step-step repulsion. The repulsive interaction between steps, controlled by an additional parameter, is introduced into the recently studied atomistic scale model of vicinal crystal growth, based on cellular automata. It is shown that the repulsion modifies bunching process in a different way, depending on the direction of the destabilizing force. In particular, bunch profiles, stability diagrams and time-scaling dependences of various bunch properties are affected when the step-step repulsion increases. The repulsion between steps creates a competition between two characteristic sizes - bunch width and macrostep height, playing the role of the second length scale that describes the step bunching phenomenon. A new characteristic time scale dependent on the step-step repulsion parameter emerges as an effect of interplay between (01) faceted macrosteps and (11) faceted bunches. The bunch height being the major characteristic size of the bunches is not influenced dramatically by the repulsion.

# 1. Introduction

The development of experimental techniques for surface analysis, technological challenges and increasing the available computational power make the surface stability studies more intense. Among several different types of instabilities at the surface step bunching (SB) is one of the most important processes. The interest in SB is closely connected with its influence on the nanostructure layer-by-layer epitaxial growth, carried out in various deposition techniques having their industrial realizations. In this context different materials are studied such, as: $CH_3NH_3PbI_3$ [1], GaN [2-4], AlN [5], AlGaN [6], SiC [7−9], graphene [10, 11], W [12, 13], $SrRuO_3$/(001) $SrTiO_3$ [14], KDP [15−19], Si [20-23], ferritin [24], and many others [25,26]. Step bunching is also observed in growth from solution, where SB development is controlled by fluid flow [27-31]. Bunching can be found not only at surfaces but in such exotic localizations as the sidewalls of the nanowires [32]. It can be shown on the base of Burton−Cabrera−Frank (BCF) [33] type of model that the directional asymmetry (bias) in the diffusion of the charged surface adatoms causes that the motion of the steps at the vicinal surface is unstable [34−36]. The phenomenon of SB was investigated actively in the following years by experimental techniques [1−29], theoretically [33−53] and by numerical simulations [54−61]. The phenomenon of destabilization by both step-down (SD) and step-up (SU) direct currents was at first observed by Latyshev et al. [20] on sublimating Si(111) surfaces. The observed destabilization for Si surface happened at different temperatures. Recently it was found that both current directions across the steps can create bunches at a given temperature. Such phenomenon was studied on the surface of tungsten W(110) [13,14], on the insulator $Al_2O_3$ [14] and recently at Si(001) [23].

Our recent studies of 1D atomistic scale model of growing vicinal surface (called shortly *vicCA*) destabilized by drift of adatoms show that the instability is induced by any of two opposite drifts in the two fundamental situations of step motion [60]. The *vicCA* model as a simple combination of cellular automaton and Monte Carlo steps realizing together system time evolution allows for study of the scaling behavior of bunching phenomenon in long runs of large systems. The fine-tuning of step transparency bound to the adatom diffusion and step kinetics was realized and studied by means of this model. We confirm the value of the numerical prefactor in the time scaling of bunch size *N*, by results obtained from systems of ordinary differential equations (ODE) for the step velocity. In contrast to *vicCA*, the ODE model in natural way contains step-step repulsion. In the current work we propose a constraint, which when added to the existing model works as a step-step repulsive force. We show how the presented picture, which includes surface profiles, stability diagrams, scaling exponents and prefactors, depends on the increasing strength of step-step repulsive interaction. It appears that the repulsion between steps has different influence onto the step bunching dynamics for SD and SU driving force. While in the SD case bunching process is present for all possible strengths of the repulsion, answering to its existence by the shape transformations, the mechanism of SU bias induced bunching is more delicate. Bunches become smaller and smaller with increasing step-step repulsion, and they decay to some negligible size for very strong repulsion. Thus the bunching process becomes suppressed when a strong repulsive force is acting between steps. Such analysis shows how different in their essence are bunching mechanisms induced by both SD and SU currents, even if the final effect looks similar for some growth parameters.

A special case of SB is the macrostep formation when the distances between the steps do not exist anymore due to a strong destabilization effect and/or weak step−step repulsion. This phenomenon was studied both experimentally and numerically [58-60]. We show that step-step repulsion does not prevent entirely the existence of macrosteps in step bunching process, except for the case when infinite repulsion between steps is applied. The macrostep creation seems to be important part of bunching process and macrostep size even plays the role of the second length scale describing the SB phenomenon with characteristic time scale proportional to $(1-P_{rep})^{-1.5}$ where $P_{rep}$ is repulsion parameter. The secondary length scale of macrostep size goes in parallel with the major length scale of bunch size. However, when step-step repulsions are introduced in the model of vicinal surface, bunch width that grows due to the emergence of (11) faceted bunches takes the role of the macrostep size as a second length scale.

A classification of the step bunching phenomena that was introduced recently [42] is based on the number of length scales necessary to describe the emerging patterns as a result of the evolving instability. The step bunching within the B1-type is a simpler case [42,62] – the bunch width and the bunch height share the same time-scaling exponent or, otherwise said, the equation connecting both is a trivial, linear dependence. Then, it is enough to determine carefully one of them [62]. In the B2-type the step bunching patterns are self-affine - there are two (perpendicular) length scales. This is a consequence of the observation that the time evolution of the bunch width $W$ is different than that of the bunch height/size (number of steps in the bunch) $N$ and it is quantified by difference in the time-scaling exponents which, in its turn, results in a set of other exponents [39]. This type of step bunching is observed when the models describing it contain in their equations of step motion terms that are different in nature and form. In such case step bunching is a sequence of emerging step-step attraction [63-65] (as a result of stress/strain accumulation during hetero-epitaxial growth) next to the omnipresent step-step repulsions (see a discussion on the effect of step-step repulsions [52,66] and the references therein, and a review on the studies of the scaling of minimal step-step distance in the bunch $l_{min}$ is available also [67]). When the pattern formation during the surface roughening is controlled by the competition between step-step repulsion and attraction, there are two possibilities – shorter-ranged repulsions [64] and shorter-ranged attractions [42,68]. While the former leads to infinite coarsening, the latter results in a length scale selection [42].

Below we show that the step bunching in our vicinal model (*vicCA*) with step-step repulsion for the case of complete, non-overlapping condition provides a counter example for the B1-type of SB and thus requires modification of the classification scheme [41] – here the time-scaling exponent of the bunch width (note that as a result of the non-overlapping condition there are no more macrosteps) and the bunch height is the same but still no step-step attraction is available in the model [69,70]. Further on we present a detailed quantitative study of the effects of short-ranged step-step repulsion in the SB phenomenon and emphasize on the essential differences in the behavior of SD and SU induced bunches in the presence of the repulsion.

The remainder of the paper is organized as follows. In Section 2 we describe briefly our model, based on cellular automata, and the applied simulation procedure. In Section 3 we present the results from our extensive simulations. We analyze the time scaling of various bunch properties under SD and SU driving forces and the stability diagram dependence on the interaction strength. Universal scaling functions are shown and discussed. We end this paper with a brief summary of the results in Section 4.

## 2. Model

To study step-step interaction influence on the step bunching process and on the bunch shape we use cellular automata model, studied before in various contexts [13,58-60,71]. It is very simple model, easy to manipulate and run, and at the same time it contains the most important elements that control bunching process. Such construction allows to study large systems in long simulation runs thus obtaining the limit necessary to find the universal rules for surface dynamics. In the present work the idea is to model the repulsion between steps by preventing the particle attachment to the step when the terrace is very narrow. If it is assumed that particles do not attach to the step when terrace has width of one lattice constant, macrosteps are not created, and the net effect is as if steps repel each other. We will generalize this idea by changing the probability for step-step repulsion $P_{rep}$ allowing for values higher than zero thus studying transition from non-interacting to the interacting term.

In general, our *vicCA* model consists of two ingredients, namely: the cellular automata (CA) part responsible for evolution of the vicinal crystal surface and Monte Carlo (MC) part representing diffusive lattice gas of atoms deposited at the surface. Surface of the model consists of steps decreasing from the left to the right and initially separated by terraces of the length $l_0$. Simulation procedure consists of following stages: first we update CA model accepting, with probability $p_G$, attachment of adatoms located on the right of the steps. The probability of acceptance depends on the length of terrace and in the basic version it is zero if the terrace length is equal to one (i.e. if the adatom attachment to step leads to the creation of macrostep). In its generalized version $p_G$ is modified by repulsion parameter $P_{rep}$ when terrace length is equal to 1. In this case the attachment probability equals $p_G(1-P_{rep})$ where $P_{rep}$ varies from 0 to 1 ($P_{rep}=0$ means no step-step repulsion, and 1 is for infinitely strong interaction). In the next stage of each time step all adatoms diffuse along the system jumping to right with probability $1/2+\delta$ and to left with probability $1/2-\delta$. The sign of $\delta$ determines the direction of applied bias thus $\delta<0$ induces step-up drift and when $\delta>0$ the drift is step-down directed. During the diffusion process all adatoms try to perform $n_{DS}$ diffusional jumps, but only those that point at an unoccupied neighboring lattice site are performed. When increasing the number of diffusional steps $n_{DS}$ one departs from the diffusion-limited (DL) growth mode toward the kinetics-limited (KL) one and, simultaneously, increases the step transparency. Next the stage of sublimation from the steps comes. The detachment rate is given by the parameter $p_S$ and it is modified by the same repulsion parameter $P_{rep}$ to the value of $p_S(1-P_{rep})$ in these cases when detachment leads to the creation of macrostep. The combination of step attachment and detachment allows to simulate growth as a reversible process. Finally, the number of particles in the adatom cloud is randomly updated by addition or removal of particles so that, at the end of each time-step, the adatom concentration equals its initial value $c_0$. Above procedure describes the sequence of a single time step and is repeated many times during each run of the simulation. Summarizing, a single time step of our vicinal model is consisted of the following four stages: (1) growth update (acceptance of adatom attachment to steps); (2) diffusion process (all adatoms perform given number of diffusional steps $n_{DS}$); (3) sublimation process (detachment from steps); (4) compensation of adatom concentration to its initial value $c_0$. Therefore it can be said that the time is measured in units of growth updates.

Below we compare the influence of step-step repulsion on the process of step bunching in system destabilized by drift in two opposite directions. The evolution of the non-interacting system in the presence of SD and SU drift was analyzed in previous studies [13,58-60]. Here we go beyond and introduce short range step-step repulsion in the system. Previous studies [13,60] show that in SU biased system the step bunching is obtained only when the growth rate is not too high. Otherwise systems do not form step bunches but roughen. Therefore, in order to ensure slow step motion and observe step bunching in the case of SU drift, one needs to make the process reversible, i.e. both processes of attachment and detachment should be present in the simulation. Thus we assume growth probability $p_G = (1-c_o+0.1)$ and sublimation probability $p_S = c_0$ in the case of growing systems with SU drift. However, system grown under SD bias is less vulnerable to the growth rate and the process can be irreversible. In such a case we set $p_G = 1$ and $p_S = 0$. Only in situations when attachment or detachment leads to the coalescence of steps, these probabilities are modified in such a way that the impact of the short-ranged repulsion between steps to be taken into account, then $p_G(1-P_{rep})$ is the probability for attachment and $p_S(1-P_{rep})$ for detachment. Thus, the repulsion prevents the steps to coalesce and therefore to create macrosteps. At infinite step-step repulsion (when $P_{rep}=1$) the modified probabilities of attachment and detachment are equal to zero and no macrostep formation can be found.

## 3. Results

We present the results from extensive computer simulations of vicinal crystal growth destabilized by step-down or step-up drift of the adatoms and show how the overall picture of the bunching process is influenced by the short range repulsion between steps. Firstly, let us provide definitions of the studied characteristic bunch properties, such as bunch size $N$, bunch width $W$, bunch slope $N/W$ and macrostep size $N_m$, which are very useful for description of the step bunching phenomenon. An important criterion defines if two neighboring steps belong to the same bunch – it is when the distance $l$ between them is less than the initial vicinal distance $l_0$. Therefore a group of consecutive steps separated by distances $l<l_0$ is considered as a *bunch*, whereas a group of coalesced steps with distance $l=0$ is referred as a *macrostep*. Bunch size $N$ measures the height interval between the topmost and the lowest steps in the bunch, whereas *bunch width W* measures the distance between them. *Bunch slope N/W* is calculated as a ratio between the height and the width of the bunch. *Macrostep size $N_m$* measures the height interval between the topmost and the lowest steps in the macrostep, whereas the width of the macrostep is always zero. When the height of the surface profile changes from site to site by one unit cell, it is understood as a single (mono) step, and when it changes by more than one unit cell, as a macrostep. In the case of no step-step repulsion incorporated into the model the resulting step bunches consist of single steps, but also of macrosteps. Therefore the macrostep appears as a part of bunch and also one bunch may contain a few macrosteps. In the case of infinite repulsion the steps repel each other and cannot coalesce together. This leads to the formation of bunches that consist only of mono-steps and macrostep formation cannot be found.

### 3.1. Consequences of short range repulsion between steps

Let us first discuss the basic case of infinitely strong, short range repulsion ($P_{rep}=1$), when steps cannot approach closer than one lattice distance and therefore cannot coalesce. In such a case behavior under influence of SD and SU bias is very different. As we will further observe, bunching proceeds in the presence of step-step repulsion but with no effect on the time dependence of bunch size *N* in system destabilized by SD bias. In the case of SU bias step-step repulsion influences bunching process in more pronounced way. Nevertheless, the repulsion affects the profiles of bunches in both bias directions.

In Fig.1 the profiles of SD current induced bunches with non-interacting (Fig.1a) and interacting (Fig.1b) steps are compared. When no repulsion is applied, $P_{rep}=0$, bunches consist mainly of macrosteps that become dominating structures visible in the surface profile and this makes the bunch slope very steep. On the contrary when an infinite repulsion is applied, $P_{rep}=1$, the surface slope becomes equal to 1 along the whole bunch (see the inset in Fig.1b). It means that bunch consists only of single steps separated by terraces of length 1. Very regular bunched structure appears. Thus the step bunching in *vicCA* model in this case provides a counter example for the mentioned earlier B1-type of SB. Here the time-scaling exponents of the bunch width and the bunch height are the same, as it will be shown further, so one length scale is present in this case. On the other hand in SU growing system strong repulsion destroys the creation of bunches, and surface profiles stay flat and rough. The reaction on the presence of step-step repulsion in both SD and SU cases is drastically different. In SD case the bunch slope becomes constant, and bunches are regular, while in SU case bunches are destroyed by strong interactions. Below, when the repulsion energy is attributed to the suppression of attachment/detachment probability at narrow terraces, we can study systematically the surface instability as a function of step-step repulsion strength.

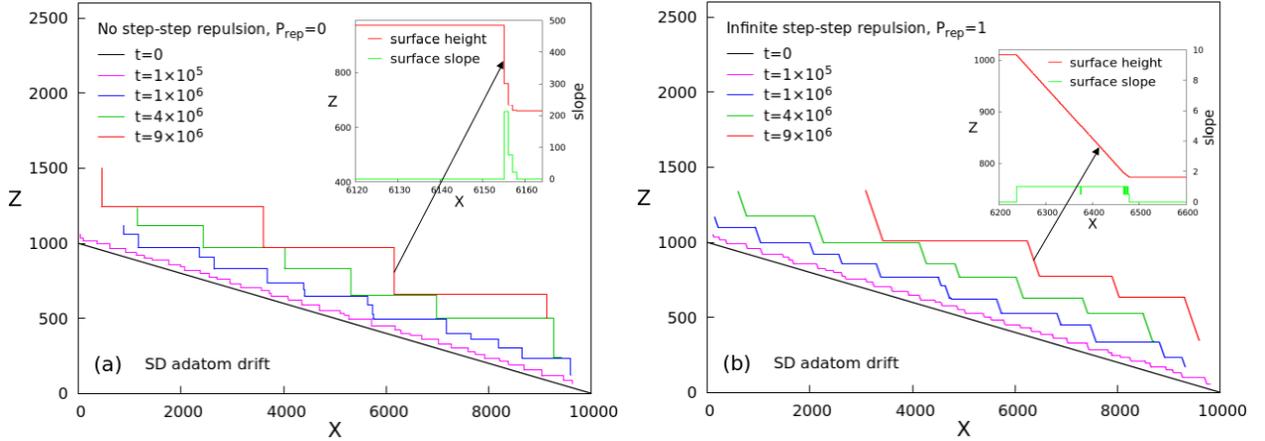

Figure 1. Time evolution of growing vicinal surface and comparison of bunch shapes in case of (a) no repulsion and (b) infinite repulsion between steps in SD biased system with parameters $l_0=10$, $c=0.1$, $\delta=0.2$, and $n_{DS}=1$.

### 3.2. Time-scaling of bunch size

We present systematic study of the step bunching sensitivity to the strength of short range step-step repulsion. Firstly, we investigate the time evolution of the bunch size $N$ for growing systems with SD and SU drift of the adatoms for different sets of system parameters. We apply various numbers of diffusional steps $n_{DS}$, biases $\delta$, adatom concentrations $c_0$, initial vicinal distances $l_0$, and step-step repulsion parameters $P_{rep}$. Bunch size $N$ is calculated as a sum of all steps in the group where each step is closer than distance $l_0$ to the next one and then we calculate the mean value of the size of all bunches in the system. It was recently [60] shown that time-scaling of the mean size of bunches $N$, induced by SD current, is not dependent on $c_0$, hence we do not investigate its impact here for SD case. To obtain proper time scaling parameters we perform calculations for large systems (surface with 1000 initial steps), large number of time steps (~$10^7$) and average the calculations over three runs for each set of parameters. The rescaled results are shown in Fig.2a and 2b for SD and SU case, respectively. Horizontal axis represents the rescaled time $T$, which is dimensionless, and denotes the real simulation time $t$ rescaled with an appropriate combination of system parameters. Fig.2a shows that introduction of the repulsion between steps in SD biased system does not affect the time dependence of the bunch size,

$N(T) \sim T^\beta$, where the scaling exponent $\beta=1/2$. It turns out that the scaling exponent and also the scaling prefactor remain unchanged when repulsion is applied in SD case. The scaling in SU case is more delicate, as one can see in the Fig.2b. It involves more parameters and the concentration $c_0$ appears in the formula for scaling prefactor. But in the place of $c_0$ we use an *effective* adatom concentration at steps defined as $c_0' = c_0 - c_0(1 - c_0 + 0.1)$ and obtain an universal curve describing bunch size behavior $N = \frac{2}{\sqrt{3}} T^{1/2}$ in both SD and SU cases with $T$ formulated as:

$$T = \frac{\delta n_{DS}}{4 l_0} t \qquad \text{for SD bias,} \quad \delta > 0 ,$$

$$T = \frac{|\delta| n_{DS}}{2 l_0} c_0'(1 - c_0')(1 - P_{rep})^{0.4} t \qquad \text{for SU bias,} \quad \delta < 0 . \qquad (1)$$

Note that the step-step interaction strength has its contribution to the time-scaling dependence of bunch size only in the case of SU bias, while in SD case time-scaling prefactor does not depend on $P_{rep}$. Nevertheless, the time-scaling exponent of bunch size, $\beta=1/2$, remains the same with increasing strength of repulsion in both cases. Note, that scaling formulas in Eq.(1) are valid only in this region of parameter space, where regular bunching happens. In the case of SU bias it is true only for low repulsion strength, $P_{rep} < 0.75$, because above this strength bunching process is destroyed. This will be discussed in more details further on. One can see that again, depending on the bias direction, step-step interaction strength differently influences step bunching process.

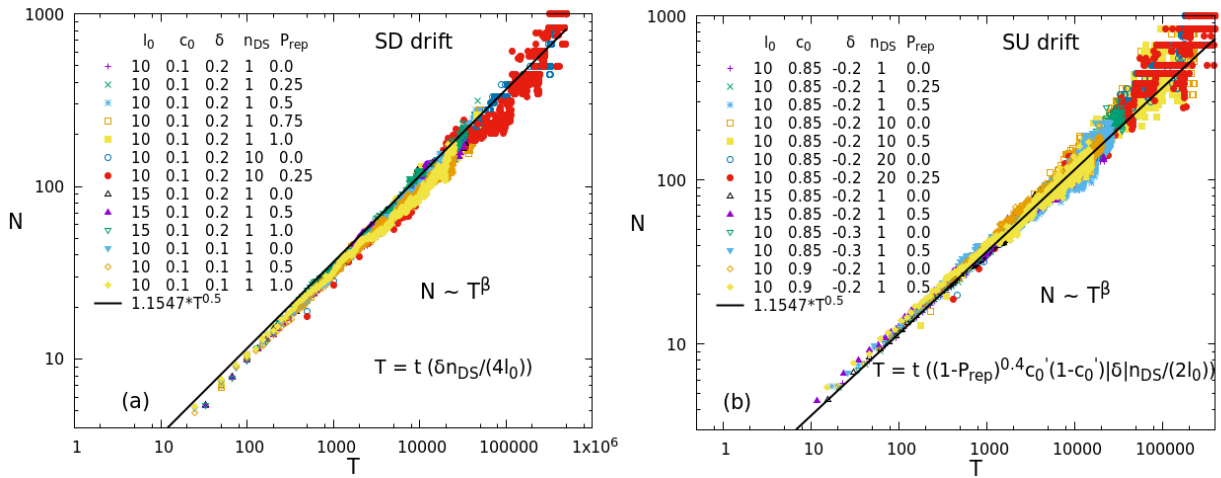

Figure 2. Time-scaling of bunch size $N$ for (a) SD and (b) SU drift of the adatoms for different parameters.

### 3.3. Stability diagrams

Time-scaling that is shown above is used to construct the stability diagrams of the model. Such diagrams give an information about the optimal values of the parameters where step bunching develops during vicinal growth and this corresponds to the most intensive bunch size $N$. Each point of the diagram, which corresponds to particular set of system parameters, is simulated for the same rescaled time $T=1000$ that in general means various numbers of time steps. Thus the length of each run depends on the coordinates in the parameter space. The obtained stability diagrams of bunch size $N$ as a function of concentation $c_0$ and interaction parameter $P_{rep}$ are shown in Fig.3 for SD and SU case. In SD biased systems (Fig.3a) SB emerges at concentration below 0.55 [60], and is present at all possible values of $P_{rep}$ (including non-interacting case). Moreover, in the whole region of system parameters where step bunching occurs, after the same period of time $T=1000$ we observe bunches with the same size $N\approx30$. Therefore both parameters, concentration and repulsion, do not affect the time-scaling dependence of bunch size in SD case, as it is already shown in Eq.(1). The maximal possible value of $c_0$ in bunching region slowly decreases at higher $P_{rep}$ thus in the infinite repulsion case its maximal value is around 0.44. Note also that at $c_0=0$ step motion is very slow or absent thus emergence of SB is not possible. It is represented in Fig.3a by dark stripe at the bottom of the parameter space. In SU biased systems (Fig.3b) step bunching is observed only for not so strong repulsions $P_{rep}<0.75$ and in rather narrow region of concentrations between 0.75 and 0.95. Obviously, in SU case high interaction strength suppresses bunching. It can be seen that for $P_{rep}<0.5$ bunch size $N$ reaches around 30 for the time $T=1000$, rescaled by the relation for SU bias in Eq.(1). For $P_{rep}$ between 0.5 and 0.75 step bunching is still present, but the size of bunches is decreased and reaches no more than 20. Finally, at very high repulsion strength, $P_{rep}>0.75$, we still observe some bunches but their size $N$ is very small and it even doesn't increase in time but saturates. Thus the step bunching process in SU biased systems does not develop but becomes suppressed when a strong repulsion ($P_{rep}>0.75$) acts between steps, which inevitably deteriorates the time scaling of bunch size and other bunch properties. It can be seen that SB is induced by both step-up and step-down external biases, but the process happens at different ranges of adatom concentrations, moreover step-step repulsive interaction influences this process in completely different way. Below we will analyze the character of SB in both cases.

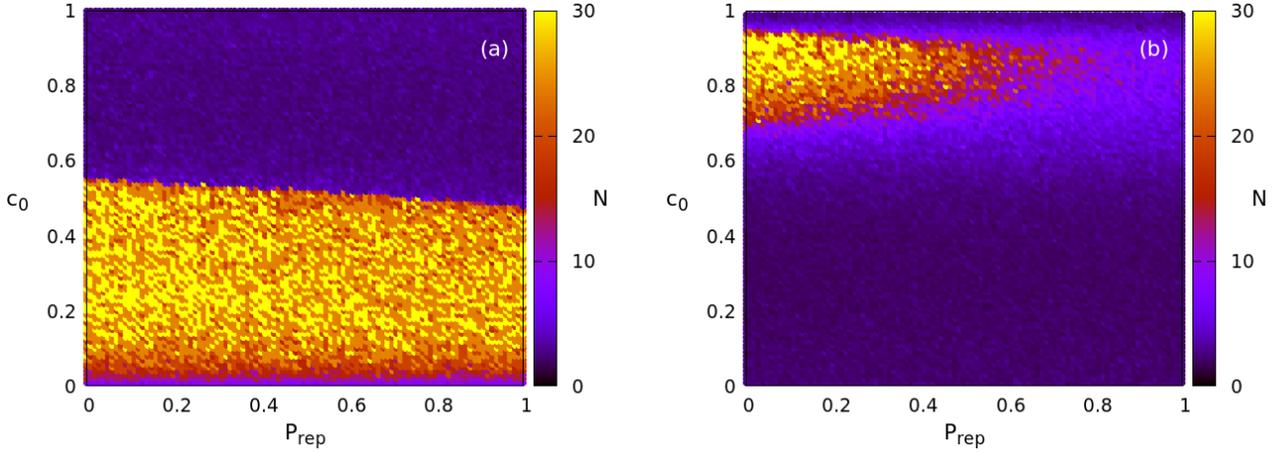

Figure 3. Stability diagrams of bunch size $N$ as a function of concentration $c_0$ and interaction parameter $P_{rep}$. Bunches are measured after rescaled time $T=1000$, where $n_{DS}=1$, $l_0=10$, and $\delta=0.2$ for SD bias (a), or $\delta=-0.2$ for SU bias (b).

### 3.4. Time-scaling of bunch width and bunch slope

Here we enhance the study of SB and investigate the time-scaling of width $W$ and slope $N/W$ of the bunches induced by SD and SU currents. Increasing systematically the strength of repulsion we study time evolution of the mean bunch width $W(t)\sim t^x$ and of the mean bunch slope $N(t)/W(t)\sim t^y$ in SD biased system (with parameters $l_0=10$, $c=0.1$, $\delta=0.2$, and $n_{DS}=1$), Fig.4a and 4c, and in SU biased system (with $l_0=10$, $c=0.85$, $\delta=-0.2$, and $n_{DS}=1$), Fig.4b and 4d. Bunch slopes and widths are averaged over all bunches observed during each time step of simulation. Thus we obtain the scaling exponents, $x$ of the bunch width and $y$ of the bunch slope, for different strengths of the repulsion. Both exponents, $x$ and $y$, are related to each other and to the value of $\beta$. Using already obtained time-scaling of the bunch height, $N(t)\sim t^\beta$ with $\beta=1/2$, one can derive the time-scaling of the bunch slope, $N(t)/W(t)\sim t^y$, where the exponent $y=\beta-x$. Although the exponent $\beta$ is constant in the whole range of $P_{rep}$ (see Fig.2), it appears that both other exponents, $x$ and $y$, depend on the repulsion parameter. One can clearly see (from Fig.4) that stronger repulsion induces bunches with greater width and smaller bunch slope, which reflects respectively in increasing value of $x$ from 0.05 to 0.25 and decreasing value of $y$ from 0.45 to 0.25 when finite repulsion is applied in both SD and SU biased systems. Furthermore, when infinite step-step repulsion ($P_{rep}=1$) is applied, bunch width and bunch height are observed to scale in time with the same exponent $x=\beta=1/2$ (for SD case only) and as a result, the bunch slope scales in time with exponent $y=\beta-x=0$ (see Fig.4c). Moreover, when $P_{rep}=1$, the mean bunch slope tends to be nearly 1 for SD case. It means that the bunch width is approximately equal to the bunch height and the resulting bunches consist of single steps separated by single lattice distances, which means that it is observed (11) faceted bunch. In SU biased systems as the repulsion gets stronger the step bunching process becomes more suppressed, respectively bunch height, bunch width, and bunch slope stop to increase in time and their scaling dependences reach saturation at some time. A possible explanation could be that the rate of surface growth becomes higher for stronger repulsions and as it is known those systems that grow at high velocity do not form bunches but roughen. That's why in the case of SU biased system with strong repulsion between steps we observe this saturation in the time-scaling dependences of all characteristic bunch properties. It is also observed that when the repulsion gets stronger the time to reach the saturation is moved towards earlier times, and obviously the data after these times could not converge to any scaling master curve (and this should be taken into account).

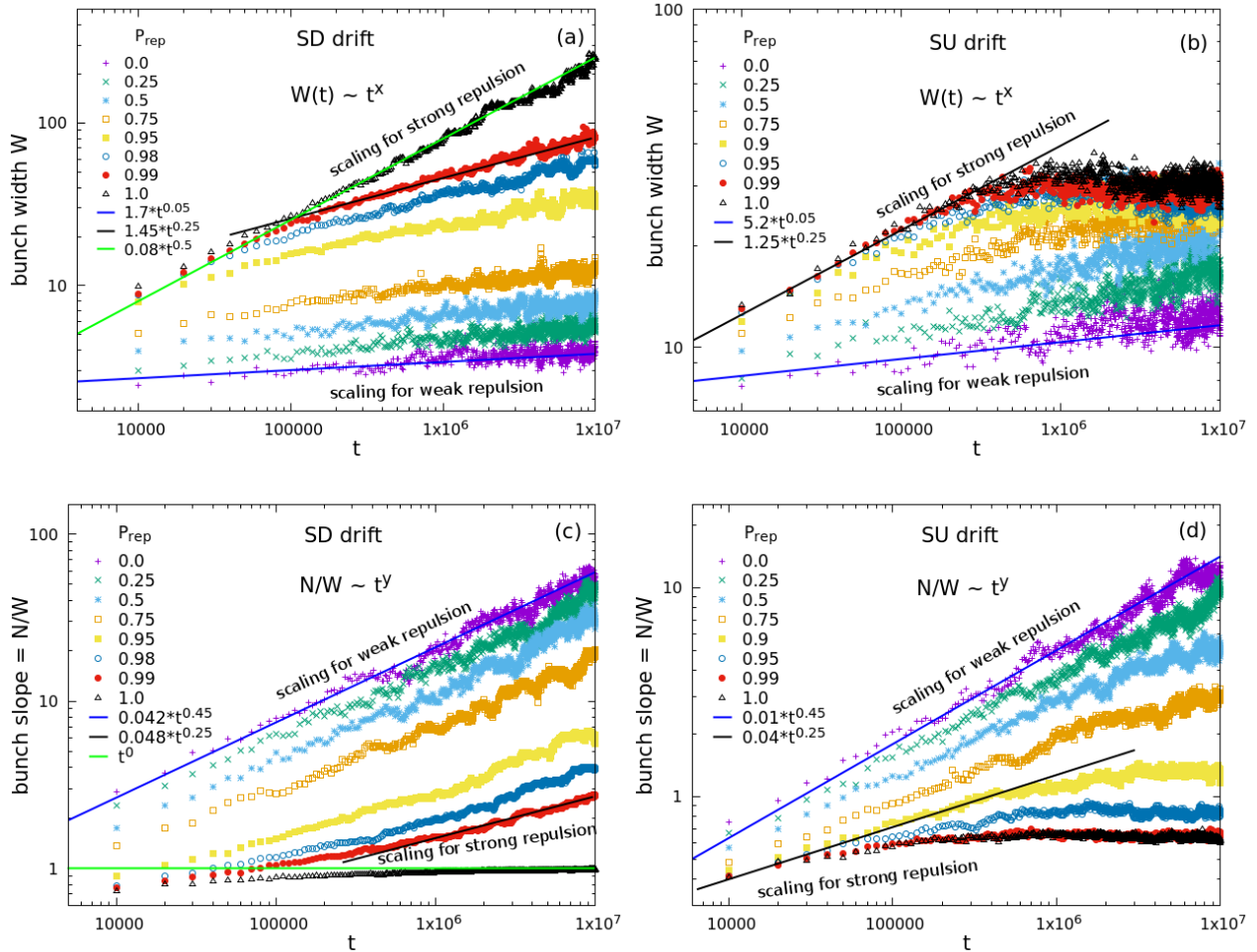

Figure 4. Time-scaling of bunch width $W$ and bunch slope $N/W$ for different step-step repulsion parameter $P_{rep}$ in SD biased system, (a) and (c), with $l_0=10$, $c=0.1$, $\delta=0.2$, and $n_{DS}=1$, and in SU biased system, (b) and (d), with $l_0=10$, $c=0.85$, $\delta=-0.2$, and $n_{DS}=1$. Bunch width $W$ versus time $t$ is shown in panels (a) and (b) and bunch slope $N/W$ versus time $t$ - in panels (c) and (d).

At first sight in Fig.4c and 4d one may conclude that the scaling exponent y of bunch slope seems to decrease gradually with increasing the strength of repulsion. However, we were able to find an appropriate rescaling with regard to the repulsion and as a result all studied curves representing bunch slope for different repulsions are collected along a single universal (master) curve. Therefore the results for bunch slope can be presented as a function of time rescaled regarding to the repulsion parameter, as it is shown in Fig.5a and 5b for SD and SU case, respectively. It appears that the universal scaling curve of bunch slope has a crossover behavior with two clearly outlined trends which correspond to two different regimes of step bunching - at weak and strong repulsion. Thus, we obtain the universal scaling exponent $y$ with two values that correspond to these two regimes: $y = 0.45$ and $0.25$ at weak and strong repulsion, accordingly, as shown in the legends of Fig.5a and 5b.

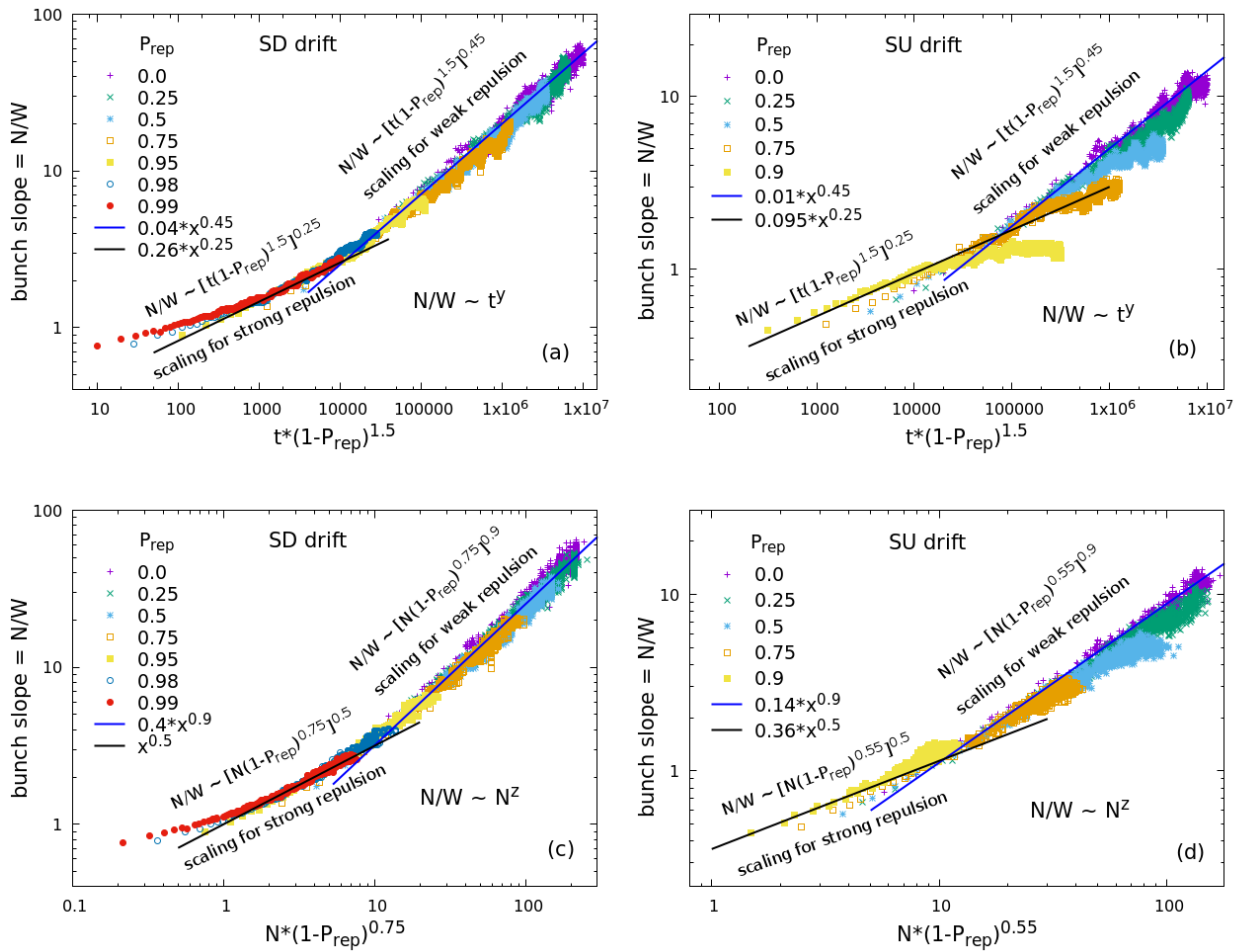

Figure 5. Scaling relations of bunch slope $N/W$ for different step-step repulsion parameter $P_{rep}$ in SD case, (a) and (c), where $l_0=10$, $c=0.1$, $\delta=0.2$, and $n_{DS}=1$, and in SU case, (b) and (d), where $l_0=10$, $c=0.85$, $\delta=-0.2$, and $n_{DS}=1$. Bunch

slope versus time *t* is shown in panels (a) and (b) and bunch slope as a function of bunch size *N* is shown in panels (c) and (d).

Moreover, it is clearly seen (in Fig.5a and 5b) that any curve for a given repulsion starts to scale with the one (lower) exponent in earlier times and then continues to scale with the other (higher) exponent in later times. We should note also that in SD case when the repulsion tends to infinity the curves of bunch slope at the beginning have a general trend towards exponent *y*=0. This corresponds to the situation when both characteristic bunch lengths, bunch width *W* and bunch size *N*, scale in time with the same exponent 0.5. Furthermore, if carefully look at the obtained scaling relations in Fig.5a and 5b, one can see that the values of the exponent *y* are the same for SD and SU case and even the scaling prefactor is the same, $(1-P_{rep})^{1.5}$, and it remains unchanged for the whole range of repulsion parameter in both cases. This corresponds to exactly the same contribution of the repulsion strength to the scaling of bunch slope in both bias directions.

Similarly, in Fig.5c and 5d we present the bunch slope *N/W* dependence on the bunch size *N* in the same systems with SD and SU bias. The mean bunch slope scales with bunch height as $N/W \sim N^z$ where the exponent *z* can be calculated as $z = y/\beta = (\beta-x)/\beta$. It appears that *z* is also dependent on the strength of repulsion parameter. The obtained values of this exponent in the corresponding universal curve are: *z* = 0.9 and 0.5 at weak and strong repulsion, respectively, and they remain the same for SD and SU case. Only the scaling prefactor is different: $(1-P_{rep})^{0.75}$ for SD current and $(1-P_{rep})^{0.55}$ for SU current, but in each current direction it remains unchanged for weak and strong repulsion regimes.

Further, it turns out that the repulsion parameter has a different contribution to the time-scaling of bunch width $W(t) \sim t^x$ at weak and strong repulsion in each bias direction, respectively $(1-P_{rep})^{13.5}$ and $(1-P_{rep})^{1.5}$ for SD bias, and $(1-P_{rep})^{9.5}$ and $(1-P_{rep})^{0.7}$ for SU bias. Because the scaling prefactor differs at weak and strong repulsion, the data for *W*(t) cannot be rescaled at once in the whole repulsion range so as to collapse onto a single master curve. It appears also that the time-scaling exponent *x* is different at weak and strong repulsion, *x* = 0.05 and 0.25 respectively. However, the obtained values of *x* are observed to be the same in SD and SU biased systems.

It should be noted that bunch size is not dependent on the repulsion in SD case, $N(t) \sim t^{0.5}$, as it is shown in Fig.2, but the situation is not the same in SU case, where $N(t) \sim [t(1-P_{rep})^{0.4}]^{0.5}$. Therefore, one should take into account these two different time-scaling dependences of the bunch size in order to achieve the consistency between all scaling relations obtained above. Finally, after detailed analysis of various bunch properties, the scaling dependences of bunch slope and bunch width can be summarized in the following Table 1:

*Table 1*

Contribution of step-step repulsion $P_{rep}$ to scaling dependences of bunch slope $N/W$ and bunch width $W$ for growing system under SD and SU drift of adatoms.

| Scaling dependences | | Bunch slope $N/W \sim t^y$ | Bunch slope $N/W \sim N^z$ | Bunch width $W \sim t^x$ |
|---|---|---|---|---|
| SD drift | weak repulsion | $N/W \sim [t(1-P_{rep})^{1.5}]^{0.45}$ | $N/W \sim [N(1-P_{rep})^{0.75}]^{0.9}$ | $W \sim [t/(1-P_{rep})^{13.5}]^{0.05}$ |
| | strong repulsion | $N/W \sim [t(1-P_{rep})^{1.5}]^{0.25}$ | $N/W \sim [N(1-P_{rep})^{0.75}]^{0.5}$ | $W \sim [t/(1-P_{rep})^{1.5}]^{0.25}$ |
| SU drift | weak repulsion | $N/W \sim [t(1-P_{rep})^{1.5}]^{0.45}$ | $N/W \sim [N(1-P_{rep})^{0.55}]^{0.9}$ | $W \sim [t/(1-P_{rep})^{9.5}]^{0.05}$ |
| | strong repulsion | $N/W \sim [t(1-P_{rep})^{1.5}]^{0.25}$ | $N/W \sim [N(1-P_{rep})^{0.55}]^{0.5}$ | $W \sim [t/(1-P_{rep})^{0.7}]^{0.25}$ |

We are able to explore the scaling dependences of the bunch height $N$, the bunch width $W$ and the bunch slope $N/W$ during the growth of vicinal surfaces under SD and SU biases at different strength of step-step repulsion. It turns out that all scaling exponents $x$, $y$, $z$ and $\beta$ are interrelated, but only $\beta$ is not dependent on the repulsion. All other exponents depend strongly on the repulsion strength. Moreover, we obtain universal scaling dependences of the studied quantities with crossover behavior where the different values of the scaling exponents are assigned to different regimes of step bunching at weak and strong repulsion.

### 3.5. Creation of macrostep and time-scaling of macrostep size

The creation of macrosteps is an important part of bunching process. When repulsion is not applied bunches consist mainly of macrosteps and also of single steps, but the former becomes dominating structure. Thus inevitably, apart from the bunch size, the macrostep size appears to be an appropriate quantity for the description of SB phenomenon. Below we demonstrate how the introduction of the repulsion between steps prevents the existence of macrosteps in the step-bunching process and how it reflects on the time-scaling of macrostep size. Recent studies [59] of systems with non-interacting steps show that for the case of biased diffusion the mean macrostep size scales in time as $N_m(t) \sim t^{\beta_m}$, where $\beta_m = 3\beta/4 = 3/8$ in the diffusion-limited (DL) regime ($n_{DS} \geq 1$) and $\beta_m = 3\beta/5 = 3/10$ in the kinetics-limited (KL) regime ($n_{DS} \gg 1$). Now we study the influence of step-step repulsion in DL regime when the number of diffusional steps $n_{DS}$ is not so big.

In Fig.6a and 6b we show how the repulsion affects the time evolution of the mean macrostep size $N_m(t)$ in SD biased system (with parameters $l_0=10$, $c=0.1$, $\delta=0.2$, and $n_{DS}=1$) and in SU biased system (with $l_0=10$, $c=0.85$, $\delta=-0.2$, and $n_{DS}=1$), respectively. One can notice that the size of the macrosteps decreases gradually with increasing repulsion and they still exist even at very strong repulsion, especially in SD case. In between the observed (10) macrosteps more and more structures of (11) orientation emerge. However in SD biased system growing with infinite repulsion between steps ($P_{rep}=1$), (10) macrosteps disappear and the resulting structure consists only of (11) faceted bunches (which contain only single steps). In contrast, when SU driving force is applied, a very strong repulsion $P_{rep}>0.75$ destroys not only macrosteps, but also the creation of bunches. However, at lower values of the repulsion between steps in SD and SU biased systems macrostep size always achieves, sooner or later, a power law dependence on time with exponent $\beta_m=3/8=0.375$. It is in agreement with recent studies [59] of the same time-scaling in DL regime with non-interacting steps. Although the time-scaling exponent $\beta_m$ of macrostep size seems to decrease gradually with increasing the strength of repulsion, we manage to find an appropriate rescaling with regard to the repulsion and as a result all studied curves representing macrostep size for various repulsions collapse onto a single universal curve. The rescaled time dependences of $N_m$ at various system parameters, including $P_{rep}$, are presented in Fig.6c and 6d for SD and SU case, respectively. The resulting universal curve for macrostep size $N_m(T) \sim T^{\beta_m}$ scales in time with the same exponent $\beta_m=3/8$ for SD and SU driving force, only the scaling prefactors are different, as follows:

$$T = \frac{\delta(1-P_{rep})^{\frac{3}{2}}}{4 l_0} t \qquad \text{for SD bias}, \quad \delta > 0,$$

$$T = \frac{\delta^2(1-P_{rep})^{\frac{3}{2}} n_{DS}^{\frac{1}{2}} c_0'(1-c_0')}{2 l_0} t \qquad \text{for SU bias}, \quad \delta < 0.$$

(2)

Since at stronger repulsions macrosteps have a tendency to disappear, then the universal power law tends to scale with zero exponent which is clearly seen in initial times.

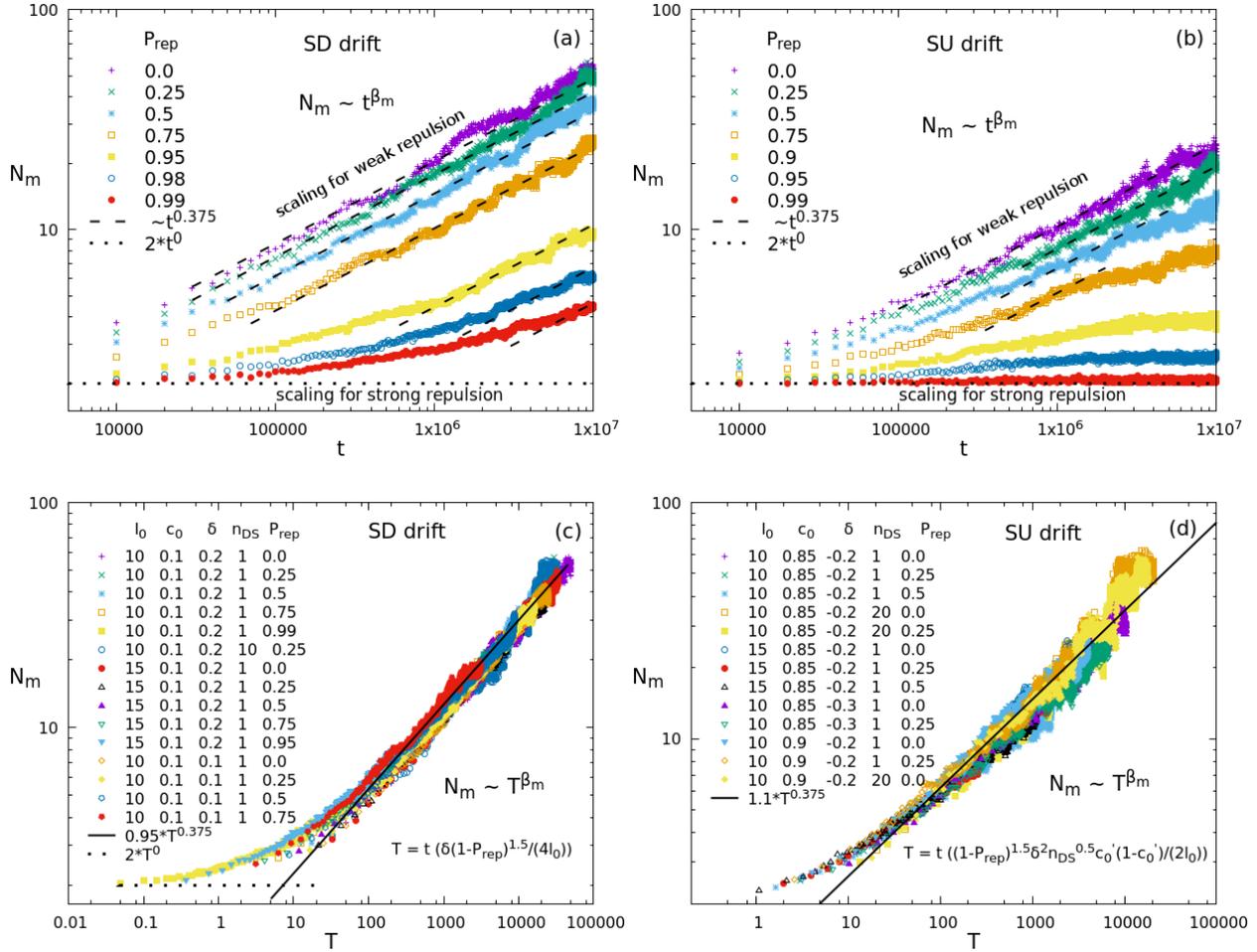

Figure 6. Time-scaling of macrostep size $N_m(t)$ for different step-step repulsion parameter $P_{rep}$ in (a) SD biased system with $l_0=10$, $c=0.1$, $\delta=0.2$, and $n_{DS}=1$, and in (b) SU biased system with $l_0=10$, $c=0.85$, $\delta=-0.2$, and $n_{DS}=1$. Time-rescaling of macrostep size $N_m(T)$ for different system parameters is shown in (c) for SD and in (d) for SU case.

Note also that the repulsion strength has the same contribution to the time-scaling dependence of macrostep size in both bias directions, $N_m(t) \sim [t(1-P_{rep})^{3/2}]^{3/8}$. It means that repulsion is related to the new characteristic time scale $\sim(1-P_{rep})^{-3/2}$ that describes the reduction of macrostep size under SD and SU drift. This time scale measures the time needed for the bunching instability to reach the asymptotic state - it becomes infinite for $P_{rep}=1$, and is finite for any other value $P_{rep}<1$. Scaling prefactors in Eq.(2) indicate that although the rest system parameters influence in a different way the time-scaling of macrostep size under SD and SU bias, the strength of step-step repulsion has the same impact on macrostep size during the bunching process in both bias directions (at least for weak repulsion strength), which is in contrast to the behavior of bunch size. In conclusion, the effect of step-step repulsion on time-scaling dependences of bunch size and macrostep size for growing systems under SD and SU drift of adatoms can be summarized in the following Table 2:

*Table 2*

Contribution of step-step repulsion $P_{rep}$ to time-scaling dependences of bunch size $N$ and macrostep size $N_m$ for growing system under SD and SU drift of adatoms.

| Scaling dependences | Bunch size $N(t) \sim t^{\beta}$ | Macrostep size $N_m(t) \sim t^{\beta_m}$ |
|---|---|---|
| SD drift | $N(t) \sim t^{0.5}$ | $N_m(t) \sim [t(1-P_{rep})^{1.5}]^{0.375}$ |
| SU drift /weak repulsion/ | $N(t) \sim [t(1-P_{rep})^{0.4}]^{0.5}$ | $N_m(t) \sim [t(1-P_{rep})^{1.5}]^{0.375}$ |

### 3.6. Bunch width versus macrostep size

Based on an atomistic-scale model we were able to qualitatively obtain step bunching in both SD and SU bias directions. Bunch size $N$ is the major characteristic length scale of step bunching which behaves as a universal one. It was shown that the time scaling exponent of the bunch size is universal and independent of the strength of step-step repulsion, only the scaling prefactor in the case of SU bias is considered slightly to depend on the repulsion. Also, it was shown previously that for the need of a second characteristic length scale, the bunch width $W$ is inevitably substituted by the behavior of the macrostep size $N_m$. And here comes the role of step-step repulsion in bunching process of crystal surfaces. We show how the step-step repulsion, introduced in our atomistic-scale model, affects the second length scale and how the macrostep withdraws from the description of SB phenomenon and again on the stage appears the bunch width. Therefore, by varying the repulsion between steps, there is a competition for the role of the second length scale between both characteristic lengths - bunch width $W$ versus macrostep size $N_m$. The time dependence of the relation between both lengths, $W/N_m$, shown in Fig.7a and 7b for

SD and SU case respectively, demonstrates precisely and quantitatively which is the dominating length. It is evident that macrostep size $N_m$ is the dominating length at weak repulsion and bunch width $W$ is at strong repulsion. As it can be seen, at the repulsion close to 0.75 one of them takes an advantage. If comparing Fig.4a and 4b with the corresponding Fig.6a and 6b, one can see that at weak repulsion $W$ does not change too much in time (being almost constant), but then $N_m$ is the important and dominating second length scale. At strong repulsion $P_{rep}>0.75$, $N_m$ does not increase so much in time and even vanishes at infinite repulsion ($P_{rep}=1$), but $W$ takes over and becomes dominating second length scale. Moreover, for (11) faceted bunches observed at infinite repulsion (for SD case) the bunch width $W$ becomes equal to the bunch height $N$ and both lengths (the first and the second characteristic bunch lengths) even scale in time with the same universal exponent ($\sim t^{0.5}$).

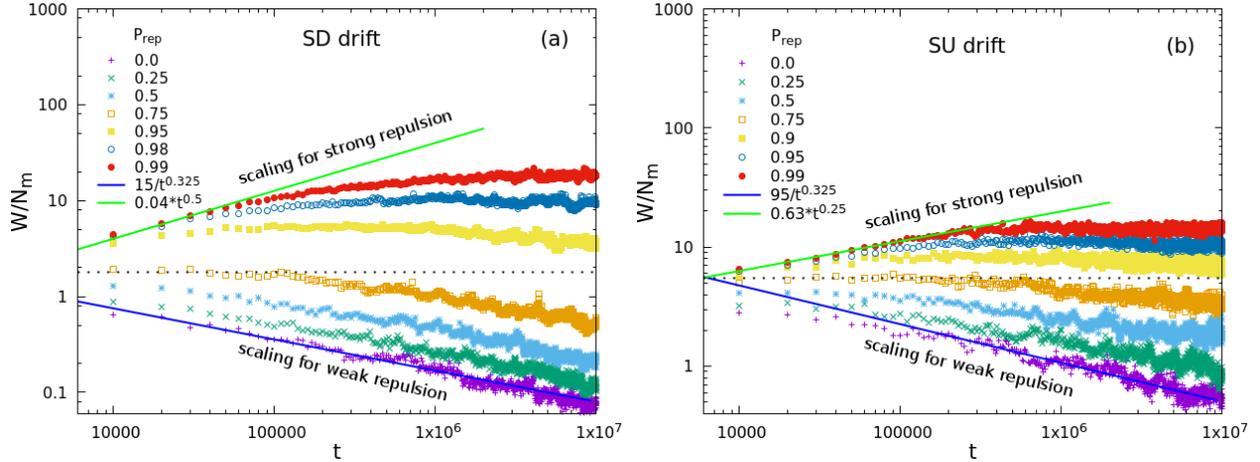

Figure 7. Time-scaling of the relation between bunch width $W$ and macrostep size $N_m$ for different step-step repulsion parameter $P_{rep}$ in (a) SD biased system with $l_0=10$, $c=0.1$, $\delta=0.2$, and $n_{DS}=1$, and in (b) SU biased system with $l_0=10$, $c=0.85$, $\delta=-0.2$, and $n_{DS}=1$.

## 4. Summary and conclusions

We have presented a quantitative study of the effect of short-ranged step-step repulsion on bunches induced by both SD and SU driving forces. Step-step repulsion is an important element of step bunching process description. Its origin is usually attributed to the surface tension and seems to be natural at each stepped surface. In the analytical approaches, based on the Burton Cabrera Frank approach, the introduction of the step-step interaction is necessary to get stable solutions. The strength and the exact form of the interaction are not clear. This is why in many approaches step bunching process is studied as a function of interaction parameters.

We have investigated the model, which allows for study of step bunching phenomenon of interacting and non-interacting steps. In our cellular automata model of vicinal crystal growth destabilized by SD or SU adatom drift bunches have finite size and the process of bunching is stable without any interaction between steps. It is still interesting how bunch properties change with increasing step-step interaction force. We have introduced repulsion to the model by additional parameter that changes particle adsorption and desorption at short, one site long terraces.

We have studied in detail the universality of the step bunching process as a function of the repulsion strength. We have shown that the interaction has different impact on the bunching process under SD and SU driving forces. While in the SD case the process of bunching continues with the same intensity for any interaction, in the case of SU bias the bunching is suppressed (destroyed) at strong repulsion between steps. At the same time the shape of bunches changes with repulsion under SD bias, becoming more flat, whereas SU biased bunches almost do not change their shape with repulsion even though their size is reduced. Moreover, when step-step repulsion is introduced in the atomistic-scale model, bunch width replaces macrostep size as a second length scale needed to describe the SB phenomenon, whereas the bunch size used as a major length scale is not influenced so much by the repulsion. It turns out that the repulsion strength has the same contribution $(1-P_{rep})^{3/2}$ to the time-scaling of macrostep size in both SD and SU bias directions. This is in contrast to the time-scaling of bunch size which is slightly affected by the repulsion only in the case of SU bias. Nevertheless, the time-scaling exponents of bunch size ($\beta=1/2$) and of macrostep size ($\beta_m=3/8$) are not changed with increasing the strength of repulsion in both SD and SU cases. New characteristic time scale $\sim(1-P_{rep})^{-3/2}$ is seen. It can be attributed to the process of reduction of macrostep size and creation of new, (11) faceted bunches. They are clearly seen in SD biased system. The time-scaling dependences of bunch width and bunch slope are also studied. The scaling of bunch slope in both bias directions is observed to have a universal crossover behavior with time-scaling exponent changing from 0.45 to 0.25 when the repulsion between steps is increased, which corresponds to two different regimes of step bunching depending on the strength of interaction - weak or strong repulsion. Nevertheless, the repulsion strength has absolutely the same contribution $(1-P_{rep})^{3/2}$ to the universal time-scaling of bunch slope in both SD and SU bias directions. However, we obtain a completely different contribution of the repulsion parameter to the time-scaling of bunch width at weak and strong repulsion in each direction. In the presented study we have highlighted the essential differences in the behavior of SD and SU induced bunches in the presence of short-ranged repulsion between steps. Different reactions of bunching process on the interaction strength for different bias would allow to speculate about the interaction from the given behavior of the experimental system under the influence of bias.

**Acknowledgments**

The present research is supported by the National Science Centre (NCN) of Poland (Grant No. 2013/11/D/ST3/02700). H.P. and V.T. acknowledges partial financial support from grant T02-8/121214 with the Bulgarian NSF. Part of the calculations was done on HPC facility Nestum (BG161PO003-1.2.05). H.P. would like to thank the Institute of Physics at the Polish Academy of Sciences for financial support and hospitality during her stay.